\documentstyle[preprint,aps]{revtex}
\input epsf.tex
\def\DESepsf(#1 width #2){\epsfxsize=#2 \epsfbox{#1}}
\begin{document}

\preprint{\vbox{\hbox{OITS-644}\hbox{COLO-HEP-394}\hbox{}}}
\draft
\title {Branching Ratios and CP Asymmetries of B Decays to a Vector and a
Pseudoscalar Meson}
\author{\bf N.G. Deshpande $^{\dagger} $, B. Dutta$^{\dagger} $, and Sechul Oh
$^{\ast} $ }\address{$^{\dagger} $ Institute of Theoretical Science, University
of Oregon, Eugene, OR 97403\\$^{\ast} $ Department of Physics, University of
Colorado, Boulder, CO 80309 }
\date{December, 1997}
\maketitle
\begin{abstract} 
 We consider two body decays of B meson into a light vector (V) and 
a pseudoscalar (P) meson. The constraint obtained from the $B\rightarrow P\,P$ 
modes on the
parameter space of the input parameters is imposed also on $B\rightarrow  V
\,P$ modes. In particular we constrain $\xi\equiv (1/N_c)$ for those modes from
recently measured $B\rightarrow \omega K,\,\phi K$ and are able to get a
satisfactory pictures for all modes where data exists. Modes that should be seen
shortly and those with possibly large CP asymmetries are identified. 
\end{abstract}

\newpage Recently CLEO \cite{cleo0,cleo1} has reported information on the 
branching ratios of a number of exclusive modes where $B$ decays into  a pair of
pseudoscalars ($P$) or  a vector ($V$) and a pseudoscalar meson. In some cases they
have seen the decay  for the first time and in other cases they have improved
the bounds. Among the
$B$  to $PP$ modes it was found that the branching ratio of the mode
$B\rightarrow
\eta^{\prime} K$ is larger than expected.  We have shown that \cite{ddo3} by a 
suitable  choice of parameters, this large branching ratio can be explained in
the context of factorization technique without invoking any new  physics or high
charm content in $\eta^{\prime}$. The same parameter space is also found to
account for other
$B\rightarrow PP$ modes (e.g. $B\rightarrow \pi K$ and $B\rightarrow\pi\pi$) 
consistent with the experimental data. The parameters involved in the above 
process are  1) effective number of color
$\xi(\equiv 1/N_c)$, 2) CKM angles and phases 3) Form factors 4) the QCD scale 
and 5) the light quark masses. In this paper we calculate all $B\rightarrow VP$
modes using the same parameter space as was used in the case of
$B\rightarrow PP$ modes, with the exception of $\xi$. This is because $\xi$ 
which  takes account of non-factorizable contributions could in principle be
different  in $VP$ final states.  We also need one new form factor which appears
in the matrix element of the states involving B meson and  a vector meson. We
use the experimental bound on 
$B\rightarrow
\omega K$ and
$B\rightarrow\phi K$ to put restriction on $\xi$ and the new form factor. We 
then predict the branching ratios and the CP asymmetries for all the other
modes,  some of which are expected to be measured soon.

The calculations proceed in two steps. First we consider the effective short
distance Hamiltonian in the next to leading order (NLO). We then use the 
generalized factorization approximation to derive hadronic matrix elements by
saturating the vacuum state in all possible ways. The effective weak Hamiltonian
for hadronic $B$ decays can be written as 
\begin{eqnarray}
 H_{\Delta B =1} &=& {4 G_{F} \over \sqrt{2}} [V_{ub}V^{*}_{uq} (c_1 O^{u}_1
+c_2 O^{u}_2) 
   + V_{cb}V^{*}_{cq} (c_1 O^{c}_1 +c_2 O^{c}_2)
   - V_{tb}V^{*}_{tq} \sum_{i=3}^{12} c_{i} O_{i}] \nonumber \\ 
  &+& h.c. ,
\end{eqnarray}  where $O_{i}$'s are defined as 
\begin{eqnarray}
 O^{f}_{1} &=& \bar q_{\alpha} \gamma_{\mu} L f_{\beta} \bar f_{\beta}
\gamma^{\mu} L b_{\alpha} ,
  \ \  O^{f}_{2} = \bar q \gamma_{\mu} L f \bar f \gamma^{\mu} L b ,          
\nonumber \\ 
 O_{3(5)} &=& \bar q \gamma_{\mu} L b \Sigma \bar q^{\prime} \gamma^{\mu} L(R)
q^{\prime} ,   
  \ \ O_{4(6)} = \bar q_{\alpha} \gamma_{\mu} L b_{\beta} \Sigma \bar
q^{\prime}_{\beta} 
       \gamma^{\mu} L(R) q^{\prime}_{\alpha}  ,                        
\nonumber \\ 
 O_{7(9)} &=& {3 \over 2} \bar q \gamma_{\mu} L b \Sigma e_{q^{\prime}} \bar
q^{\prime} 
       \gamma^{\mu} R(L) q^{\prime} ,        
  \ \ O_{8(10)} ={3 \over 2} \bar q_{\alpha} \gamma_{\mu} L b_{\beta} \Sigma
e_{q^{\prime}} 
       \bar q^{\prime}_{\beta} \gamma^{\mu} R(L) q^{\prime}_{\alpha} \;,
\end{eqnarray}  where $L(R) = (1 \mp \gamma_5)/2$, $f$ can be $u$ or $c$ quark,
$q$ can be $d$ or $s$ quark,  and $q^{\prime}$ is summed over $u$, $d$, $s$, and
$c$ quarks.  $\alpha$ and $\beta$ are  the color indices, $c_i$s  are the Wilson
coefficients (WCs). $O_{7-10}$ are the electroweak penguin  operators due to
$\gamma$ and $Z$ exchange, and the ``box'' diagrams at loop level. We shall
ignore the smaller contributions from the dipole penguin operators. The initial
values of the WCs are derived from the matching condition at the $m_W$ scale.
However we need to renormalize them \cite{A} when we use these coefficients at 
the scale
$m_{b}$. We will use the effective values of WCs at the scale
$m_b$ from the ref
\cite{ddo3}. We have shown that in $B\rightarrow PP$ case there is very little 
$\mu$  dependence in the final states. 

The generalized factorizable approximation has been quite successfully used in
two body $D$ decays as well as $B\rightarrow D$ decays \cite{stech0}. The method 
includes color octet non factorizable contribution by treating $\xi\equiv 1/N_c$
as an adjustable parameter \cite{desh,neubert,stech2}. Other work related to our
paper  includes  ref.\cite{{AG},{CT}} who have considered $B\rightarrow \omega
K$ and 
$B\rightarrow \omega\pi$. These results are consistent with ours when restricted 
to our choice of parameter space. Work also exists on decays purely based on 
$SU(3)$ symmetry \cite{ros}. This approach though more general, lacks detailed 
predictions that we can make using generalized factorization. Let us now 
describe the parametrization of the matrix elements and the decay constants in
the case of
$B\rightarrow VP$ decays \cite{AG}.
\begin{eqnarray}
\langle
M(p^{\prime})|V_{\mu}|B(p)\rangle&=&[(p+p^{\prime})_{\mu}-{{m_B^2-m_M^2}\over
q^2}q_{\mu}]F_1(q^2)+{{m_B^2-m_M^2}\over q^2}q_{\mu}]F_0(q^2)\\
\langle V(\epsilon, p^{\prime})|(V_{\mu}-A_{\mu})|B(p)\rangle&=&{2\over
{m_B+m_V}}i\epsilon^{^*}p^{\alpha}p^{\prime\alpha}V(q^2)\\\nonumber
&-&(m_B+m_V)[\epsilon_\mu-{\epsilon^*.q\over q^2}q_{\mu}]A_1(q^2)\\\nonumber
&+&{{\epsilon^*.q}\over{m_B+m_V}}[(p+p^{\prime})_{\mu}-{{m_B^2-m_V^2}\over
q^2}q_{\mu}]A_2(q^2)\\\nonumber &-&\epsilon^*.q{{2m_V}\over q^2}q_{\mu}A_0(q^2)
\end{eqnarray} and the decay constants are given by:
\begin{eqnarray}
\langle 0|A_{\mu}|M(p)\rangle &=& i f_{M} p_{\mu},\,
\langle 0|V_{\mu}|V(\epsilon,p)\rangle =i f_{V} m_{V}\epsilon_{\mu},
\end{eqnarray} where $M$, $V$, $V_{\mu}$ and $A_{\mu}$ denote a pseudoscalar
meson, vector meson, a vector current and an  axial-vector current,
respectively, and
$q=p-p^{\prime}$.   Note that $F_1(0)=F_0(0)$ and we can set 
$F^{B\rightarrow M}_{0,1}(q^2=m^2_M)
\approx F^{B\rightarrow M}_{0,1}(0)$ since these form factors are pole dominated
by mesons at scale $m^2_B$. Among all the form factors in the $\langle 
V(\epsilon, p^{\prime})|(V_{\mu}-A_{\mu})|B(p)\rangle$ matrix element, only 
$A_0$ survives when we calculate the full $B\rightarrow VP$ decay amplitude.
The $A_0$ is related to $A_1$ and $A_2$:
\begin{eqnarray}  A_0(0)&=&{{m_B+m_V}\over {2m_V}}A_1(0)-{{m_B-m_V}\over
{2m_V}}A_2(0).
\end{eqnarray} The form factors are related to each other by flavor SU(3)
symmetry. For a current of the type $\bar u\gamma_{\mu}(1-\gamma_5)b$ we have 
following  values of $A_0$ for $B$ decaying into $K^*,\, \omega \, {\rm and}\,
\rho$
\begin{eqnarray}  A_0^{B\rightarrow\omega}&=&{G\over \sqrt
2},\,A_0^{B\rightarrow K^*}=G,\, A_0^{B\rightarrow\rho}={G\over \sqrt
2}.\nonumber
\end{eqnarray} The values of the form factors present in the decay amplitude of
$B$ into pseudoscalars are:
\begin{eqnarray}  F_{0,1}^{B\rightarrow K}&=&F,\,F_{0,1}^{B\rightarrow
\pi^{\pm}}=F,\, F_{0,1}^{B\rightarrow \pi^{0}}={F\over\sqrt 2},\\\nonumber
F_{0,1}^{B\rightarrow\eta^{\prime}}&=&F({{\rm sin\theta}\over\sqrt 6}+{{\rm
cos\theta}\over\sqrt 3}),\,F_{0,1}^{B\rightarrow\eta}=F({{\rm
cos\theta}\over\sqrt 6}-{{\rm sin\theta}\over\sqrt 3}).\nonumber
\end{eqnarray} In ref.\cite{ddo3}, we find $F=$0.36 gives a good fit to
$B\rightarrow PP$ data and we shall find that $G=$0.28 in  ref.\cite{AG,PB} 
provides a good fit to
$B\rightarrow VP$ decays. The values of the decay constants (in MeV) we use 
are\cite{stech0,AG,PB,PT}:
\begin{eqnarray}  f_{\omega}&=&195,\,f_{K^{*}}=214,\, 
f_{\rho}=210,\,f_{\pi}=134,\, f_K=158,\, f_1=f_{\pi},\,f_8=1.75 f_{\pi}.\nonumber
\end{eqnarray} The decay constant $f^{u,s}_{\eta^{\prime}}$ and $f^{u,s}_{\eta}$
are obtained by combining $f_1$ and $f_8$ with a $\eta-\eta^{\prime}$ mixing
angle
$\theta$, where $\theta$ is $-25^{0}$. The particular choices $f_1$ and $f_8$
and 
$\theta$ value allow us to fit the $B\rightarrow \eta^{\prime}K$ experimental
bound without  violating any other experimental constraints. In  processes
involving $\eta^{\prime}$, when estimating we have included the effects of
anomaly \cite{AG,PT,AC}.

For the preferred value of $\gamma$, we use ref. \cite{ddo3} where the ratio of 
the branching ratio of
$B\rightarrow\eta^{\prime} K$ and the branching ratio of
$B\rightarrow\pi K$ has been studied. The ratio does not depend on the form
factors and it has been  found that the small value of the weak phase 
$\gamma\simeq 35^0$ is preferred. It has also been pointed out that in order to
satisfy the  experimental constraint on the branching ratio of $B\rightarrow
\pi^+\pi^-$ mode, the smaller
$|V_{ub}/V_{cb}|=0.07$ is preferred.

Now we discuss the constraints on $\xi$ that can be obtained from the branching
ratio  of
$B\rightarrow \phi K$ and $B\rightarrow \omega K$. Recent measurement at CLEO
\cite{cleo1} yield the following  bounds:
\begin{eqnarray} BR(B^\pm \rightarrow \omega K^\pm)= (1.5^{+0.7}_{-0.6}\pm
0.3)\times 10^{-5},
\\\nonumber
 BR(B^{\pm} \rightarrow \phi K^{\pm})< 0.53 \times 10^{-5}. \nonumber
\end{eqnarray} In the figure 1 we have plotted the branching ratio of $B^\pm
\rightarrow \omega K^\pm$ averaged over particle-antiparticle decays as a
function of $\xi$ for $\mu=m_b$. We calculate the branching ratio by multiplying
the partial width of the particular mode by the total rate
$\tau_B=1.49$ ps. We can see from the figure that only  the large values of
$\xi>0.6$ or small values of $\xi<0.15$ are experimentally allowed. For most of
the $\xi$ values,  the penguin part of the amplitude is larger than the tree
part by almost an order of magnitude. The region of
$\xi$ (0.3 - 0.5) where the branching ratio is smallest, the penguin part and the
tree part are of the same order. Here the CP asymmetry is also very large ($\sim
69\%$).  The asymmetry is however small (at the most
$4\%$)  for the $\xi$ values which are allowed by the experiment.  The branching
ratio is not sensitive to the form factor
$A_0$. 

In the figure 2 we plot the  branching ratio $B^\pm \rightarrow \phi K^\pm$
averaged over particle-antiparticle decays as a function of $\xi$ for $\mu=m_b$.
From the figure we see that only the lower values of $\xi(<.3)$ are allowed. In
this mode there is no tree contribution, and also does not depend on $A_0$.
Considering this B decay mode and the mode discussed above, we can conclude that
only the lower values of
$\xi\leq 0.2$ are allowed by the $B\rightarrow VP$ decays.

In the figure 3 we show the branching ratio $B^\pm \rightarrow \omega
\pi^\pm$ averaged over the  particle-antiparticle decays as a function of $\xi$
for $\mu=m_b$. The tree part is larger than the penguin part by an order of
magnitude. There exists a CLEO observation \cite{cleo1} ($(1.1^{+0.6}_{-0.5}\pm 
0.2)\times 10^{-5}$) for this  mode, but the experimental result has large
errors. When the result  improve this mode will be a crucial test for
factorization hypothesis. As it stands now there is mild disagreement with data
that prefers larger
$\xi$ values. This rate can be enhanced somewhat by larger $G$. We show this as
an example for $G=$0.35 in the figure 3 by a dashed line. For the rest of the 
decays we have used $G=$0.28.

The branching ratios of other $B\rightarrow VP$ modes, we have calculated, are
all smaller than the experimental bound currently available. In Table 1 we have 
shown the average branching ratios of all the charged $B$ decay modes to a vector
and a pseudoscalar for $|\Delta S|=0\, {\rm and}\,1$, where $S$ is the 
strangeness quantum number. In Table 2, we shown the branching ratios of the
neutral $B$ decays to a vector and a pseudoscalar and Table 3 we have shown the
available experimental bounds \cite{cleo1} at 90 $\%$ C.L. except those entries
with  errors which are the observed branching ratios. 

Now we discuss some of the modes whose BR are close to the recently obtained
CLEO data. We also discuss the CP asymmetries of these modes. It appears from
Table 1 and Table 2 that some  of the
$B\rightarrow\rho\pi$ modes and
$B\rightarrow K^* \pi$  modes will be observed soon.  The mode
$B^0\rightarrow
\rho^{+}\pi^-$ has  the tree part larger than  the penguin part in the amplitude
by one order of  magnitude for any value of $\xi$. The BR decreases from the
maximum value of 
$3\times 10^{-5}$ calculated at $\xi=$0, as $\xi$ increases. The asymmetry is 
$\sim$4$\%$ over the full range of $\xi$. The experimental bound on this mode 
at 90 $\%$ C.L. is  8.8$\times 10^{-5}$. The mode $B^0\rightarrow \rho^{-}\pi^+$ 
(not ${\bar B^0}\rightarrow \rho^{-}\pi^+$) has the tree part  larger than  the
penguin part in the amplitude by three order of magnitude for  any value of
$\xi$. The BR decreases from the maximum value of
$6.4\times  10^{-5}$ calculated at $\xi=$0, as $\xi$ increases. The asymmetry is
less than  1$\%$ over the full range of $\xi$. The experimental bound on this
mode at 90 $\%$ C.L. is  8.8$\times 10^{-5}$. Note that, for both those modes 
asymmetry measurement will need tagging.  The mode
$B^+\rightarrow\rho^0
\pi^+$ has the  penguin part and the tree part of  the same order for the $\xi$
values between 0 and 0.2. For $\xi$s larger than  that, the tree part is larger
than the penguin part by an order of magnitude.  The asymmetry varies between 0
to 29 $\%$, the maximum occurs at $\xi=0$. The BR  increases as $\xi$ increases.
The experimental bound at 90 $\%$ C.L. is 5.8 $\times 10^{-5}$,  which is little
larger than the largest theoretical BR 1.4$\times 10^{-5}$ which occurs at 
$\xi=$1. The mode
$B^0\rightarrow K^{*+} \pi^-$  has the penguin part larger than  the tree part
in the amplitude by one order of  magnitude for any value of $\xi$. The BR
decreases from the maximum value of 
$1.43\times 10^{-5}$ calculated at $\xi=$0  as $\xi$ increases. The asymmetry 
varies between 5 to 7$\%$ over the full range of $\xi$. The experimental bound
on  this mode at 90 $\%$ C.L. is $6.7\times 10^{-5}$. The mode
$B^+\rightarrow K^{*0}\pi^+$ is also a pure penguin process.  The BR decreases
from the maximum value of $1\times 10^{-5}$ (at $\xi$=0) as 
$\xi$ increases. The experimental bound on the BR on this mode at 90 $\%$ C.L.
is $3.9 \times  10^{-5}$. The mode $B^0\rightarrow  K^{*0} \pi^0$ has the
penguin part larger than  the tree part in the amplitude  by two order of
magnitude for lower values of $\xi$ ($\xi<0.4$). For larger values of $\xi$, the
tree part and the penguin part become comparable and gives rise to large CP
asymmetry.  The BR from the maximum value  of
$3.89\times 10^{-6}$ calculated at
$\xi=$0 decreases as
$\xi$ increases. The  asymmetry is between 0 to 30$\%$ over the full range of
$\xi$. The experimental  bound on this mode at 90 $\%$ C.L. is $2\times
10^{-5}$.  The mode
$B^+\rightarrow K^{*+}\pi^0$ has the penguin part in the  amplitude to be larger
by an order of magnitude than the tree part for almost  any value of $\xi$. The
BR decreases from the maximum value of $8.6\times  10^{-6}$ (at $\xi$=0) as
$\xi$ increases. The asymmetry is $\sim 5\%$ for any  value of $\xi$. The
experimental bound on  this mode at 90 $\%$ C.L. is $8\times 10^{-5}$.

Apart from the above discussed modes there exists some more modes, where large
CP asymmetry can be observed. For example, $B^+\rightarrow\rho^+\eta$ mode has
as large as 35 $\%$ CP asymmetry at $\xi=0$ and $\sim$ 33$\%$  at $\xi=0.2$. The
BR is expected to be $5.40\times 10^{-6}$ at $\xi=0$ and 
$5.88\times 10^{-6}$  at $\xi=0.2$.  The mode
$B^+\rightarrow\rho^+\eta^{\prime}$  has
 large  CP asymmetry $\sim$31 $\%$  at $\xi=0$ and at $\xi=0.2$. The BR is
expected to be 
$3.19\times 10^{-6}$ at $\xi=0$ and 
$2.99\times 10^{-6}$  at $\xi=0.2$. There are no experimental bounds on the
above modes yet. The mode
$B^0\rightarrow\rho^0\pi^0$  has
 CP asymmetry $\sim$ 11 $\%$  at $\xi=0$ and $\sim$ 24 $\%$  at $\xi=0.2$. The 
BR calculated at $\xi=0$ is $1.44\times 10^{-6}$ and at $\xi=0.2$ is $1.81\times 
10^{-7}$. The experimental limit is $1.8\times 10^{-5}$.

In conclusion, we have calculated branching ratio and CP asymmetry of all  the
$B\rightarrow VP$ decay modes. We find that if we keep $\xi<0.2$ and use the 
constraint obtained from the $B\rightarrow P\,P$ modes on the parameter space of
the input parameters, we can  fit the recently  obtained experimental bounds on
the branching ratios of  the $B\rightarrow VP$ modes. We have pointed out that
some of the modes  show  large  CP asymmetry in that region of $\xi$ and the
branching ratios of some of the  modes are also very close to the present bounds
and can be expected to be   observed soon. This will  definitely determine 
conclusively the applicability of the factorization  technique to these modes,
may also help to establish the CP asymmetries.

{\bf Acknowledgements} \\ We would like to acknowledge K.T. Mahanthappa and Jim
Smith for helpful  discussions.  This work was supported in part by the US
Department of Energy Grants No.  DE-FG06-854ER-40224 and DE-FG03-95ER40894.

\newpage
\noindent {\bf Table Caption:}\newline
\noindent {\bf Table 1}: The branching ratios and the asymmetries of all the 
charged B decay modes into a vector and a pseudoscalar meson for $\xi$=0 and 
0.2 are shown.\newline
\noindent {\bf Table 2}: The branching ratios and the asymmetries of all the 
neutral B decay modes into a vector and a pseudoscalar meson for $\xi$=0 and 
0.2 are  shown.\newline
\noindent {\bf Table 3}: The bounds on the branching ratios at 90 $\%$ C.L. on
various modes, except for those entries with error, which are the observed
modes.\newline
\noindent {\bf Figure Captions:}\newline
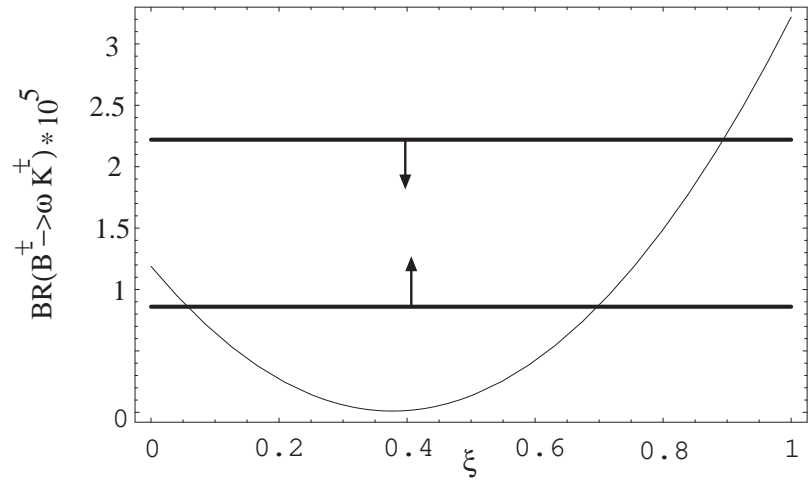
\noindent {\bf Fig. 1}: Branching ratio for the average of
$B^{\pm}\rightarrow\omega K^{\pm}$ as a function of $\xi$. The curve is for
$A_0^{B\rightarrow\omega}={0.28\over{\sqrt 2}}$.\newline
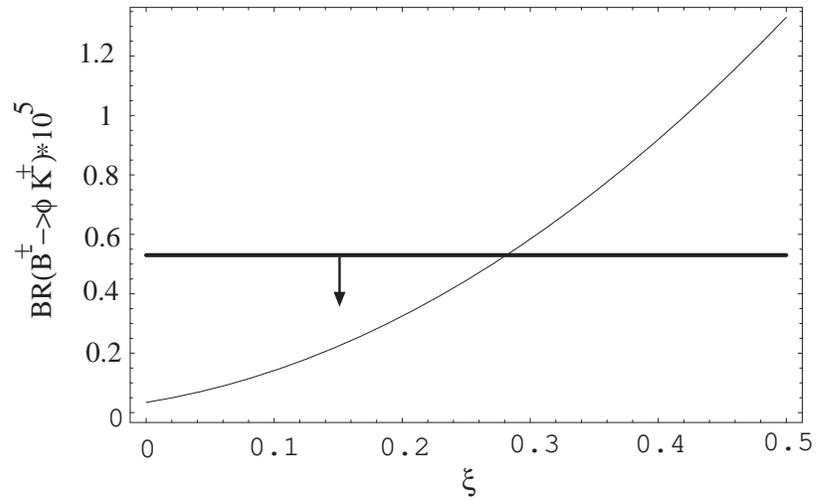
\noindent {\bf Fig. 2}: Branching ratio for the average of
$B^{\pm}\rightarrow\phi K^{\pm}$ as a function of $\xi$.\newline
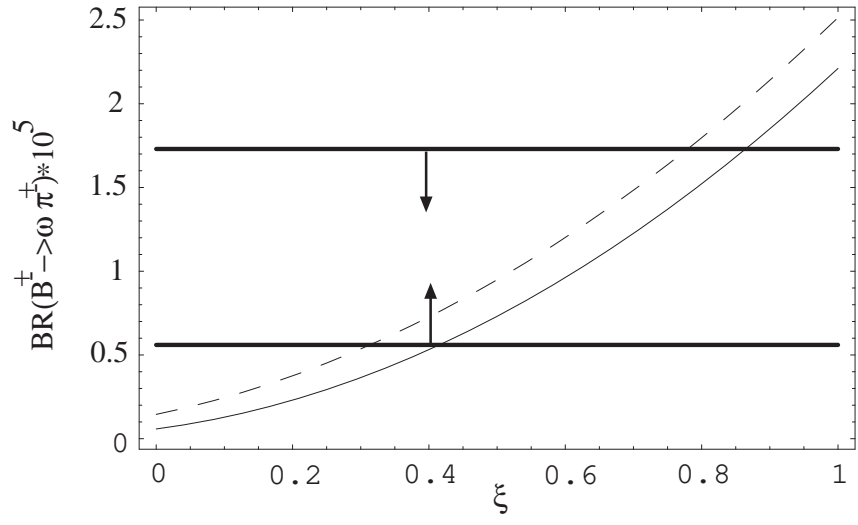
\noindent {\bf Fig. 3}: Branching ratio for the average of
$B^{\pm}\rightarrow\omega \pi^{\pm}$ as a function of $\xi$. The solid line is
drawn for $A_0^{B\rightarrow\omega}={0.28\over{\sqrt 2}}$ and the dashed line is
drawn for $A_0^{B\rightarrow\omega}={0.35\over{\sqrt 2}}$. Note that the statistical
 significance for observing this mode is only 2.9 $\sigma$.
\newpage
\begin{center}  Table 1 \end{center}
\begin{center}
\begin{tabular}{|c|c|c|c|c|c|}\hline &modes&BR &BR &asymmetry&asymmetry\\
    &  & at $\xi$=0& at $\xi=0.2$ &at $\xi$=0& at $\xi=0.2$\\\hline
 &$B^+\rightarrow\phi K^+$ &$3.42\times 10^{-7}$&$3.24\times  10^{-6}$&$
0\%$&$0\%$\\\cline{2-6}    &$B^+\rightarrow\omega K^+$ &$1.18\times
10^{-5}$&$2.71\times 10^{-6}$&$\sim  2\%$&$\sim 4\%$\\\cline{2-6}   
&$B^+\rightarrow\rho^0 K^+$ &$1.95\times 10^{-7}$&$1.18\times 10^{-7}$&$\sim 
-4\%$&$\sim -9\%$\\\cline{2-6} &$B^+\rightarrow\rho^+ K^0$ &$2.39\times
10^{-8}$&$6.43\times 10^{-8}$&$0\%$&$0\%$\\\cline{2-6} 
$|\Delta S|$=1&$B^+\rightarrow K^{*0} \pi^+$ &$9.96\times 10^{-6}$&$7.96\times 
10^{-6}$&$0\%$&$0\%$\\\cline{2-6}  &$B^+\rightarrow K^{*+} \pi^0$ &$8.59\times
10^{-6}$&$8.19\times 10^{-6}$&$\sim  5\%$&$\sim 5\%$\\\cline{2-6}
&$B^+\rightarrow K^{*+} \eta^{\prime}$ &$1.20\times 10^{-6}$&$1.50\times 
10^{-6}$&$\sim -2\%$&$\sim -2\%$\\\cline{2-6} &$B^+\rightarrow K^{*+} \eta$
&$4.09\times 10^{-6}$&$3.62\times 10^{-6}$&$\sim -7\%$&$\sim -8\%$\\\hline 
&$B^+\rightarrow\omega \pi^+$ &$5.81\times 10^{-7}$&$2.3\times 10^{-6}$&$\sim 
-5\%$&$\sim 5\%$\\\cline{2-6}  &$B^+\rightarrow\rho^0 \pi^+$ &$4.26\times
10^{-7}$&$1.62\times 10^{-6}$&$\sim -29\%$&$\sim -13\%$\\\cline{2-6} 
$\Delta S$=0&$B^+\rightarrow\rho^+ \pi^0$ &$1.16\times 10^{-5}$&$1.26\times 
10^{-5}$&$\sim 4\%$&$\sim 4\%$\\\cline{2-6} &$B^+\rightarrow\rho^+
\eta^{\prime}$ &$3.19\times 10^{-6}$&$2.99\times  10^{-6}$&$\sim -31\%$&$\sim
-31\%$\\\cline{2-6} &$B^+\rightarrow\rho^+ \eta$ &$5.40\times
10^{-6}$&$5.88\times 10^{-6}$&$\sim -35\%$&$\sim -33\%$\\\cline{2-6} 
&$B^+\rightarrow K^{*0} K^+$ &$3.52\times 10^{-7}$&$2.81\times 10^{-7}$&$ 
0\%$&$0\%$\\\cline{2-6} &$B^+\rightarrow K^{*+} K^0$ &$5.16\times
10^{-10}$&$1.66\times 10^{-9}$&$0\%$&$0\%$\\\cline{2-6} &$B^+\rightarrow\phi
\pi^+$ &$3.75\times 10^{-7}$&$8.73\times  10^{-8}$&$0\%$&$0\%$\\\hline   
\end{tabular}
\end{center}
\newpage
\begin{center}  Table 2 \end{center}
\begin{center}
\begin{tabular}{|c|c|c|c|c|c|}\hline &modes&BR &BR &asymmetry&asymmetry\\
    &  & at $\xi$=0& at $\xi=0.2$ &at $\xi$=0& at $\xi=0.2$\\\hline
&$B^0\rightarrow\phi K^0$ &$3.42\times 10^{-7}$&$3.24\times  10^{-6}$&$
0\%$&$0\%$\\\cline{2-6} &$B^0\rightarrow\omega K^0$ &$8.6\times
10^{-6}$&$1.49\times 10^{-6}$&$\sim  -2\%$&$\sim -1\%$\\\cline{2-6}    
&$B^0\rightarrow\rho^0 K^0$ &$8.21\times 10^{-7}$&$3.94\times 10^{-7}$&$\sim 
1\%$&$\sim 0\%$\\\cline{2-6}  &$B^0\rightarrow\rho^- K^+$ &$9.47\times
10^{-7}$&$8.33\times 10^{-7}$&$\sim  -4\%$&$\sim -4\%$\\\cline{2-6}  
$|\Delta S|$=1&$B^0\rightarrow K^{*0} \pi^0$ &$3.89\times 10^{-6}$&$2.84\times 
10^{-6}$&$\sim 2\%$&$\sim 1\%$\\\cline{2-6} &$B^0\rightarrow K^{*+} \pi^-$
&$1.43\times 10^{-5}$&$1.30\times 10^{-5}$&$\sim  6\%$&$\sim 6\%$\\\cline{2-6}
&$B^0\rightarrow K^{*0} \eta^{\prime}$ &$4.12\times 10^{-7}$&$8.22\times 
10^{-7}$&$\sim 1\%$&$\sim 0\%$\\\cline{2-6} &$B^0\rightarrow K^{*0} \eta$
&$8.86\times 10^{-6}$&$5.55\times 10^{-6}$&$\sim  1\%$&$\sim 0\%$\\\hline 
&$B^0\rightarrow\rho^+ \pi^-$ &$3.04\times 10^{-5}$&$2.72\times 10^{-5}$&$\sim 
4\%$&$\sim 4\%$\\\cline{2-6}  &$B^0\rightarrow\rho^- \pi^+$ &$6.38\times
10^{-6}$&$5.69\times 10^{-6}$&$\sim  1\%$&$\sim 1\%$\\\cline{2-6}
&$B^0\rightarrow\rho^0 \pi^0$ &$1.44\times 10^{-6}$&$1.81\times 10^{-7}$&$\sim 
11\%$&$\sim 24\%$\\\cline{2-6}   &$B^0\rightarrow\rho^0 \eta^{\prime}$
&$3.55\times 10^{-6}$&$2.48\times  10^{-6}$&$\sim 11\%$&$\sim 4\%$\\\cline{2-6} 
$\Delta S$=0&$B^0\rightarrow\rho^0 \eta$ &$6.68\times 10^{-6}$&$3.78\times 
10^{-6}$&$\sim 8\%$&$\sim 4\%$\\\cline{2-6} &$B^0\rightarrow\omega \pi^0$
&$1.53\times 10^{-7}$&$9.6\times 10^{-9}$&$\sim  3\%$&$\sim -65\%$\\\cline{2-6}  
&$B^0\rightarrow\omega \eta^{\prime}$ &$3.56\times 10^{-6}$&$2.49\times 
10^{-6}$&$\sim 5\%$&$\sim 2\%$\\\cline{2-6}  &$B^0\rightarrow\omega \eta$
&$7.09\times 10^{-6}$&$3.94\times 10^{-6}$&$\sim  8\%$&$\sim 4\%$\\\cline{2-6} 
&$B^0\rightarrow K^{*0} \bar K^0$ &$3.52\times 10^{-7}$&$2.81\times 10^{-7}$&$ 
0\%$&$0\%$\\\cline{2-6} &$B^0\rightarrow \bar K^{*0} K^0$ &$5.16\times
10^{-10}$&$1.66\times 10^{-9}$&$  0\%$&$0\%$\\\cline{2-6} &$B^0\rightarrow\phi
\eta^{\prime}$ &$4.13\times 10^{-8}$&$9.62\times 
10^{-9}$&$0\%$&$0\%$\\\cline{2-6}  &$B^0\rightarrow\phi
\eta$ &$1.37\times 10^{-7}$&$3.18\times  10^{-8}$&$0\%$&$0\%$\\\cline{2-6}
&$B^0\rightarrow\phi \pi^0$ &$1.87\times 10^{-7}$&$4.36\times 
10^{-8}$&$0\%$&$0\%$\\\hline 
\end{tabular}
\end{center}
\newpage
\begin{center}  Table 3 \end{center}
\begin{center}
\begin{tabular}{|c|c|c|}\hline & modes& Experimental BR ($\times 10^{-5}$) 
\\\hline &$B^+\rightarrow\phi K^+$ & $< \; 0.53$\\\cline{2-3}
&$B^0\rightarrow\phi K^0$ & $< \; 4.2$\\\cline{2-3} &$B^+\rightarrow\omega K^+$
& $1.5^{+0.7}_{-0.6}\pm 0.3$\\\cline{2-3} &$B^+\rightarrow\rho^0 K^+$ & $< \;
1.4$\\\cline{2-3} &$B^0\rightarrow\rho^{-} K^+$ & $< \; 3.3$\\\cline{2-3}
$|\Delta S|$=1&$B^+\rightarrow\rho^+ K^0$ & $< \; 6.4$\\\cline{2-3}
&$B^0\rightarrow\rho^0 K^0$ & $< \; 3.0$\\\cline{2-3} &$B^0\rightarrow K^{*0}
\eta^{\prime}$ & $< \;4.2$ \\\cline{2-3} &$B^0\rightarrow K^{*0} \eta$ & $< \;
3.3$ \\\cline{2-3} &$B^+\rightarrow K^{*+} \eta^{\prime}$ & $< \;29.0$\\\cline{2-3}
&$B^+\rightarrow K^{*+} \eta$ & $< \;20.4$\\\cline{2-3} &$B^0\rightarrow K^{*+}
\pi^-$ & $< \; 6.7$ \\\cline{2-3} &$B^0\rightarrow K^{*0} \pi^0$ & $< \; 2.0$
\\\cline{2-3} &$B^+\rightarrow K^{*+} \pi^0$ & $< \; 8.0$\\\cline{2-3}
&$B^+\rightarrow K^{*0} \pi^+$ & $< \; 3.9$\\\hline &$B^+\rightarrow\rho^0
\pi^+$ & $< \; 5.8$ \\\cline{2-3} &$B^0\rightarrow\rho^{\pm} \pi^{\mp}$ & $< \;
8.8$ \\\cline{2-3}
$\Delta S$=0&$B^0\rightarrow\rho^{0} \pi^{0}$ & $< \; 1.8$ \\\cline{2-3}
&$B^+\rightarrow\omega \pi^+$ & $1.1^{+0.6}_{-0.5}\pm 0.2$\\\cline{2-3}
&$B^+\rightarrow\phi \pi^+$ & $< \; 0.56$\\\cline{2-3} &$B^0\rightarrow\phi
\pi^0$ & $< \; 0.65$\\\hline
\end{tabular}
\end{center}
\newpage

\centerline{ \DESepsf(finalfig12bpv.epsf width 12 cm) }

\centerline{ \DESepsf(finalfig3bpv.epsf width 12 cm) }


\newpage
\begin{thebibliography}{[001]}
\bibitem{cleo0}J. Smith (CLEO collaboration), talk presented at the 1997 Aspen
winter conference on Particle Physics, Aspen, Colorado, 1997. 

\bibitem{cleo1}M.S. Alam et al (CLEO collaboration), CLEO CONF 97-23; A. 
Anastassov et al (CLEO collaboration), CLEO CONF 97-24.
 
\bibitem{ddo3}N.G. Deshpande, B. Dutta, and S. Oh, hep-ph/9710354

\bibitem{A} A. J. Buras, M. Jamin, M. E. Lautenbacher and P. Weisz, Nucl. Phys.
{\bf B400}  (1993) 37; A. J. Buras, M. Jamin and M. E. Lautenbacher, ibid (1993)
75, M. Ciuchini, E. Franco, G. Martinelli and L. Reina, Nucl. Phys. {\bf B415} 
(1994) 403. 

\bibitem{stech0}M. Neubert and B. Stech, hep-ph/9705292, to appear in Heavy
Flavors, Second Edition, ed. A. J. Buras and M. Linder (World Scientific,
Singapore).

\bibitem{desh} N. G. Deshpande, M. Gronau and D. Sutherland, Phys. Lett. {\bf
B90} (1980) 431.

\bibitem{neubert} M. Neubert (CERN), Talk presented at the International
Euroconference on Quantum Chromodynamics: QCD 97: 25th Anniversary of QCD,
Montpellier, France,  3-9 Jul 1997, hep-ph/9707368.
  
\bibitem{stech2}B. Stech, Plenary talk given at 20th Anniversary Symposium:
Twenty Beautiful Years of Bottom Physics, Chicago, IL, 29 Jun - 2 Jul 1997,
hep-ph/9709280.

\bibitem{AG} A. Ali and C. Greub,  hep-ph/9707251.

\bibitem{CT} H-Y. Cheng and B. Tseng, hep-ph/9708211.

\bibitem{ros} A. S. Dighe, M. Gronau, J.L. Rosner, hep-ph/9709223 
 
\bibitem{PB} P. Ball and V.M. Braun, Phys. Rev. {\bf D55} (1997) 5561.
\bibitem{PT} P. Ball, J.M. Frere and M. Tytgat, Phys. Lett. {\bf B365},
(1996),367. 
\bibitem{AC} A. Ali, J. Chay, C. Greub and P. Ko,  hep-ph/9712372.
\end{thebibliography}
\end{document}